\documentclass[journal]{IEEEtran}

\usepackage{cite}
\usepackage{graphicx}
\usepackage{subfigure}
\usepackage{amssymb,amsmath} 
\usepackage{multirow}

\begin{document}

\title{Data analysis with R in an experimental physics environment}

\author{Andreas Pfeiffer and Maria Grazia Pia% <-this % stops a space
\thanks{Manuscript received 15 November 2013.}% <-this % stops a space
\thanks{A. Pfeiffer is with CERN, CH-1211, Geneva, Switzerland (e-mail: Andreas.Pfeiffer@cern.ch).}
\thanks{M. G. Pia is  with INFN Sezione di Genova, Via Dodecaneso 33, I-16146 Genova, Italy 
	(e-mail: MariaGrazia.Pia@ge.infn.it).}
}

\maketitle

\pagestyle{empty}
\thispagestyle{empty}

\begin{abstract}
A software package has been developed to bridge 
the R analysis model with the
conceptual analysis environment typical of radiation physics experiments.
The new package has been used in the context of a project for the validation of
simulation models, where it has demonstrated its capability to satisfy typical
requirements pertinent to the problem domain.
\end{abstract}
%\begin{keywords}
%Monte Carlo, simulation, Geant4, X-rays
%\end{keywords}

% -----------------------------------------------------------------------------------------

\section{Introduction}
\label{sec_intro}
\IEEEPARstart{S}{oftware} tools for data analysis are widely used throughout the
experimental life-cycle in particle and nuclear physics, astrophysics and bio-medical
physics.
Among the analysis tools in these experimental environments some common
open source ones are GnuPlot\cite{gnuplot}, SciPy \cite{scipy-onl}\cite{scipy-pap}, 
ROOT \cite{root} and systems implementing AIDA (Abstract Interfaces for Data Analysis) 
\cite{AIDA} interfaces; commercial products such as MATLAB and Origin are also used.

The development described in this paper addresses the requirements of data
analysis in typical experimental physics environment through the cooperation of
two systems, which are characterized by different underlying conceptual models
and provide complementary functionality: AIDA-compliant tools and R \cite{R}.

% -----------------------------------------------------------------------------------------

\section{The AIDA  project}
\label{sec_aida}

The AIDA project started in
1999 as a collaborative effort of several developers of analysis toolkits and
frameworks who were aiming at providing a full set of abstract interfaces for
this task. 

In addition to static 1D, 2D and 3D Histograms with pre-defined
binning, a ``dynamic'' version of these histograms was defined, where the data was
stored ``as is'' for a given number of fill calls to the histogram and the binning
would be defined ``a posteriori'' given on the content (and excluding outliers) -
a first in the field at the time. 
More classical data types like Profile Histograms (also in their 1D, 2D and 3D 
incarnations), ntuples, and free-form ``Data Points'', vectors of N-dimensional 
data with errors attached, are part of the definition of the AIDA data types as 
well. 
In order to be able to attach some meta-data to the data types, an ``Annotation'' 
type was defined, allowing to add statistics and summary type of information to 
the data object as well as ``free form'' information provided by the user, e.g. 
to provide tags or labels to the object, in the form of key/value pairs of strings. 
Other interfaces were defined to describe higher level objects used in typical 
data analysis environments like Fitter, Plotter, Analyzer (``filters on the ntuple''), 
generic Functions (used for example in the Fitter) and a set of interfaces to manage 
the objects.

All the interfaces were defined for the C++ \cite{iAIDA, OpSci}, Java \cite{JAS}
and Python \cite{paida} programming languages in order to not limit applications
and users to only one language.
The main aim was to have a modular, flexible and inter-operable system using 
dynamically loadable libraries for each component, allowing the user to pick and 
choose components according to his/her requirements. This was done using the 
Factory pattern for the creation of the various objects and using plug-in modules 
for the different implementations (e.g. for different storage formats). 

In addition to defining these interfaces, an XML
format was defined for data storage and interchange, so that the files written
in this format by one application/component could be easily read by another one
implementing this standard - adding to the flexibility of the system. 

The
modularity inherent in the design of the AIDA interfaces was the reason they
were adopted in several large software projects like the Geant4 detector
simulation software \cite{g4nim,g4tns} and the Gaudi framework \cite{gaudi}, which is
used in the Atlas and LHCb experiments at the Large Hadron Collider (LHC) at
CERN, in several experiments at SLAC and in a number of other systems.

%% -----------------------------------------------------------------------------------------

% -----------------------------------------------------------------------------------------

\section{R and its analysis model}

R is a language and an open source software environment for data analysis. It
provides rich functionality for data management, statistical computation and
graphics; R kernel is complemented by a large number of specialised add-on
packages contributed by the community.

R operates on an ample variety of UNIX platforms, Windows and MacOS. 
For computationally-intensive tasks, C, C++ and Fortran code can be linked and called at run time.

R is used in a wide variety of multi-disciplinary applications; despite its
widespread use in the academic environment, it is hardly known and marginally
used in experimental environments such as 
particle and nuclear physics experiments, astrophysics research and bio-medical physics.

The use of R in these experimental environments is hindered by its underlying
analysis model: a typical  R analysis scenario assumes that the data are all available
at once, while in typical experimental environments analysis tools encompass functionality 
to produce and accumulate data managed by analysis objects in the course of a cyclic execution:
for instance, in the course of online data taking, of the generation of simulated events, of
the reconstruction and further processing of detector raw data and of physics analysis.

% -----------------------------------------------------------------------------------------

\section{Interfacing AIDA with R}
\label{sec_aidar}

Based on the well defined XML format for AIDA object storage, a module -- named \textit{aidar} -- 
to read these files into R was developed \cite{aidar}. 

This way, the user can exploit the huge power of R for analysis while using the
flexibility of AIDA in the code to generate histograms or ntuples in the
simulation or analysis of raw experimental data.

The \textit{aidar} package exploits the existing XML package \cite{XMLr} in R to
read in and parse the file and then identify the various objects in the file by
their name and convert them into R's \textit{data.frames}.

A small collection of helper functions allow the user to extract general
information on the content of the file, and to get and show annotations of
selected objects.

At the time of writing this contribution to the conference proceedings \textit{aidar} is available as a 
development version directly from the development repository.
Distribution as a regular R package is foreseen in the near future.

% -----------------------------------------------------------------------------------------

\section{Experience of use}

A set of tests of \textit{aidar} has been performed in the context of a project
for the validation of Geant4 Compton scattering simulation documented in these
proceedings \cite{nss_compton}.
Histograms and ntuples associated with physics observables relevant to the
validation process were produced in a typical Geant4 test environment,
using the iAIDA concrete implementation of AIDA interfaces.
The resulting analysis objects stored by iAIDA in XML format were imported
into R by means of the \textit{aidar} package.
The subsequent data analysis has taken place entirely in the R environment; a
sample of the results is documented in \cite{nss_compton}.

% -----------------------------------------------------------------------------------------

% -----------------------------------------------------------------------------------------

\section{Conclusion and outlook}

The \textit{aidar} package has been developed to bridge the conceptual environment
of production of data analysis objects in the course of processing a large number of events,
typical of physics experiments, with the rich functionality of R.
It has proven its capabilities in a real-life application environment, related to the validation 
of Geant4 simulation models.

The provision of the \textit{aidar} package as a regular R package is planned in a short time 
scale. 
Further tests to evaluate its performance in typical experimental environments are foreseen.

% --------------------------------------------------------------------------

% ------------------------------------------------------------------------------
%\section*{Acknowledgment}
%The authors express their gratitude to CERN for support to the research
%described in this paper.

%The authors thank ?? for proofreading the manuscript and valuable comments.

% ------------------------------------------------------------------------

\end{document}